# Monolithic Integrations of Slanted Silicon Nanostructures on 3D Microstructures and Their Application to Surface Enhanced Raman Spectroscopy

Zhida Xu[1,2*], Jing Jiang[1,2], Manas Ranjan Gartia[1,3], Gang Logan Liu[1,2]

[1]Micro and Nanotechnology Laboratory, University of Illinois at Urbana-Champaign, Urbana, IL 61801, USA

[2]Department of Electrical and Computer Engineering, University of Illinois at Urbana-Champaign, Urbana, IL 61801, USA

[3]Department of Nuclear, Plasma, and Radiology Engineering, University of Illinois at Urbana-Champaign, Urbana, IL 61801, USA

*Corresponding author:

Email: zhidaxu1@illinois.edu

We demonstrated fabrication of black silicon with slanted nanocone array on both planar and 3D micro and meso scale structures produced by a high-throughput lithography-free oblique-angle plasma etching process. Nanocones with gradual change in height were created on the same piece of silicon. The relation between the slanted angle of nanocones and incident angle of directional plasma is experimentally investigated. In order to demonstrate the monolithic integration of nanostructures on micro and meso scale non-planar surfaces, nanocone forest is fabricated on non-planar silicon surfaces in various morphologies such as silicon atomic force microscopy (AFM) tips and pyramidal pits. By integrating nanocones on inverse silicon micro-pyramid array devices, we further improved the surface enhanced Raman scattering (SERS) enhancement property of this optimized commercial SERS substrate by several folds even when using 66% less noble metal coating. We investigated the length gradient dependence and asymmetric properties of SERS effects for slanted nanocone with polarized excitation. This versatile and angle-controllable nanocone fabrication and monolithic 3D nano-micro-meso integration method provides new dimensions for production and optimization of SERS and other nanophotonic sensors.

## Introduction

Black silicon is a semiconductor material whose surface is modified with micro or nanostructures to become highly absorptive, thus appears black. It was discovered in the 1980s as an unexpected side effect of reactive ion etching (RIE) in semiconductor industry.[1] Over the years, its potential in photovoltaic antireflection layer, high sensitive photodetector, superhydrophobicity and biomedical sensing has been recognized, and hence black silicon has been produced on purpose.[2-6] Besides RIE, other methods to produce black silicon include chemical wet etching,[7,8] laser pulse irradiation,[9-11] and nanoparticle-catalyzed etch.[12] Among these techniques, RIE has the advantages of high throughput and low cost; so it is still the most widely used method. Previously, we produced black silicon with the combination of hydrogen bromide (HBr) and oxygen plasma and have demonstrated its applications in biomedical sensing and solar cell.[5,13] However in all previous cases the nanostructured black silicon were created on smooth and planar substrates and the angles of the silicon nanocones could not be controlled.

In order to produce sophisticated nanostructures such as 3D photonic crystal, angle-controllable engineering in micro and nanoscale fabrication are pursued with different methods. Oblique directional RIE with Faraday cage was developed in 1980s and has been used for producing photonic crystal.[14-16] Angle controlled ion-sputtering and focused ion-beam erosion are also used for creating nanopatterns.[17-19] The most prevalent method of producing slanted nanostructure is oblique angle deposition, or glancing angle deposition (GLAD).[20,21] Self-organized nanorod array can be produced with oblique angle deposition and has been used as humidity sensor[22], surface enhanced Raman spectroscopy (SERS) substrates[23,24], optical fiber sensor[25], plasmonic oligomer sensors[26], 2D-3D photonic crystal[27] and microbattery[28]. Similar to GLAD, the slanted nanocone black silicon (SNBS) fabrication process is a mask-free and self-organized process. The GLAD is growth process while SNBS is an etching process, which offers better cost-effectiveness and more suitable for monolithic integration. In addition, as the silicon nanocones are a part of the bulk silicon substrate, the interfacial material incompatibility issues are avoided.

In this article, for the first time, we have created black silicon with slanted nanocones, produced by a 3-step plasma passivation and etching process. The slanted angle of the nanocones can be controlled by the oblique angle of etching plasma relative to the silicon plane, even though they do not obey a linear relation. We experimentally investigated the relation between the cone slanted angle and etching angle. We have also demonstrated the fabrication of nanocones on 3D non-planar silicon surfaces such as silicon atomic force microscopy (AFM) tips and microscale pyramids. Finally, to exemplify the application of slanted black silicon, we investigated the surface-enhanced Raman scattering (SERS) properties of SNBS with different slanted angles as well as that of SNBS made on 3D non-planar surfaces, using benzenethiol monolayer as the analyte. The SERS enhancement of commercial SERS substrate structure was improved by more than 4 times with 2/3 less gold being used after the monolithic integration of the SNBS nanostructures and the enhancement factor is controllable by the geometry of SNBS. Compared to previous literatures about inorganic nanocone SERS substrates produced by etching,[29-32] there are two novelties in this work. Firstly the slanted angle of nanocone is controllable by etching plasma direction. Secondly the nanocones can be integrated on existing 3D microstructure to further enhance SERS. We also envision significant potential of SNBS in a variety of applications such as photovoltaics, biosensing and photocatalytic chemistry.

**Experimental Methods**

The black silicon is produced by a lithography-less self-masked plasma etching process. The self-mask is the dispersed oxide on surface of silicon formed by oxygen plasma. The random oxide mask protects the silicon underneath it from being etched by plasma and the nanocone array is created in this way. We have already demonstrated vertical black silicon produced by RIE and its applications in biosensing and photovoltaics.[5,13] We call it vertical black silicon because its antireflective structure is upright nanocones, sculptured on planar silicon surface by normally incident plasma.

To demonstrate that the nanocones can be produced on 3D non-planar silicon surface, especially on existing microstructures, we chose three kinds of silicon microstructures including positive pyramids, inverse pyramidal pits and sharp AFM tip. Scanning electron microscopy (SEM) images of these microscale surfaces before (insets of Figure 1a,c,e) and after (Figure 1b,d,f) the monolithic integration of slanted silicon nanocones are shown in Figure 1. The positive microscale pyramids are on the surface of commercial solar cells, produced by KOH anisotropic etching of silicon (Figure 1a, b). The inverse pyramidal pits are on surface of Klarite SERS substrate (Renishaw), produced by photolithography and KOH anisotropic etching (Figure 1c,d). Figure 1e, f show nanocones formed on an AFM silicon cantilever tip. All these surfaces turn black after the nanocones are formed on the 3D microstructures. The insets on the upper right corners of Figure 1b, d, f show the comparison of the appearances of these surfaces before and after being treated by our plasma etching process. We give the inverse silicon pyramids with slanted silicon nanocones the name black Klarite. Both the positive pyramids and pyramidal pits are created by anisotropic chemical etching of silicon (100) plane so both have exposed (111) planes with the angle of 54.7$^\circ$ with respect to the horizontal plane. [33] In this case, the angle of the incident plasma with the normal of the wall of pyramids is also 54.7$^\circ$. For the AFM silicon cantilever tip in Figure 1e,f , we notice that nanocones are formed on most surfaces only except of those sidewalls which are too steep. In Figure 1e, the sidewalls of the long sharp spike are almost vertical and they are smooth without any nanocones. But on the tip of the spike that is a bit flat, the nanocones are formed. This inspired us to investigate how the silicon nanocones can be formed on surfaces with different slanted angles.

To investigate how the nanocone are formed with different etching angle, we tilted the planar silicon wafer to a certain angle relative to the incident plasma, hoping it to be sculptured in that tilted way. Figure 2a is a schematic showing the setup of slanted etching with RIE. (1 and 7 in Figure 2a) Electrodes to create an electric field (3) meant to accelerate ions (2) towards the surface of the titled silicon sample (4). One side (right side in this diagram) of the piece of silicon (4) is blocked up by stack of glass slides and the other side (left side in this diagram) is blocked with one glass slide. The thickness of silicon piece is exaggerated for illustration. The titled angle α is determined and controlled by the height of the stack of glass slide H and the distance between the two glass stacks L. tan(α)=H/L. Figure 2b is a photograph to show how is the silicon piece mounted on carrier wafer.

In our previous work, the black silicon was produced by a one-step HBr-$O_2$ process.[5] Even though this one step process is fast, taking only a few minutes, it is not very controllable and stable. In this article, the slanted black silicon was produced with our improved RIE process, a three-step $O_2$-$CHF_3$-$Cl_2$ process at room temperature which takes less than 20 minutes in total. We have verified that this three step process is reliable and controllable. Figure 2c-e is the cross-sectional schematic of the 3-step fabrication process of slanted nanocone black silicon. In the first step, a thin film of oxide is formed on silicon surface by $O_2$ plasma (Figure 2c), this step takes 5 minutes. In the second step, $O_2$ is shut down and $CHF_3$ is flowed in for two minutes. This short period of $CHF_3$ plasma is for etching the thin oxide layer to form dispersed islands rather than for completely removing the oxide. Figure S1 in the supporting information, shows the SEM image of the dispersed oxide islands as the nanomask formed in the second step. In the third step, $CHF_3$ is shut down and the mixture of $Cl_2$ and Ar with the ratio of 10 to 1 is flowed in. This step is to etch the silicon to sculpture the nanocones with the nanomask of the oxide islands formed in step 2. $Cl_2$ is the etching gas while Ar is to boost the etching rate by physical bombardment of the silicon surface. We tried pure $Cl_2$ without Ar but the etched rate was rather slow and the silicon did not turn black. Step 3 is the highly controllable because the etching rate is determined by the radio frequency (RF) power and the gas flow rate. Under a certain etching rate, the etching depth is controlled by the etching time. In this article, we stick to the recipe with the etching rate of about 30 nm/min and the etching time of 10 minutes to get nanocones with length about 300 nm.

Even though it is a three-step process, all the three steps are carried out sequentially in the same reaction chamber and at room temperature. Therefore it is still a one step process in terms of maneuverability. Compared with our previous 1-step $HBr-O_2$ process[5], even though this 3-step $O_2-CHF_3-Ar+Cl_2$ process is a bit more complex and time-consuming, it is more controllable. In the 1-step $HBr-O_2$ process, HBr and $O_2$ are mixed thus the formation of oxide mask formation and the etching of nanocones are simultaneous.[13] Thereby different processes are entangled; it is rather difficult to quantitatively control each individual process. In this 3-step $O_2-CHF_3-Ar+Cl_2$ process, the first two steps is formation of oxide nanomask and the third step is for etching, each step is separate thus can be precisely tuned individually.

**Results**

Etching angle dependence of slanted nanocones is demonstrated and investigated. By stacking glass slides and changing the distance between the two stacks of glass slides (Figure 2a), we can set the tilted etching angle α of silicon to a certain value. The glass slides are assembled and mounted onto the sapphire carrier wafer (Figure 2b).

Figure 3 a-i are cross-sectional SEM images of SNBS after RIE treatment under different etching angles. We can see that these nanocones are about 300 nm to 400 nm in length. Some granular substances seen in Figure 3d come from the sputtered gold to avoid charging during SEM imaging. The cone slanted angle β is defined as the angle between the normal of silicon plane and the cone. Figure 3a shows the SNBS after vertical etching without tilting ($0^o$ vertical etching means that the incident plasma flow is vertical to silicon). This is the common vertical nanocone black silicon demonstrated in our previous work.[5,13] The etching angles in Figure 3a-i are $0^o$, $8^o$, $15^o$, $20^o$, $30^o$, $40^o$, $50^o$, $60^o$, $70^o$ respectively. As the etching angle α increases, the cone slanted angle β will also increase. But β is always smaller than α. When the etching angle α goes above $80^o$, the nanocones will not form and thus the silicon substrate surface will not turn black. That explains why we did not obtain nanocone structures on the nearly vertical sidewall of AFM tip in Figure 1e, f. For every SEM image in Figure 3, we measured and marked the complementary angle of cone slanted angle β. The plot in Figure 3j shows the more explicit relationship between the etching angle α and cone slanted angle β.

We already demonstrated that the nanocone slanted angle is dependent on the etching angle. A photograph of two pieces of SNBS is shown in Figure 4a. Each piece of slanted nanocone black silicon is not uniformly black. The piece shown on the left side of Figure 4a is the same one shown in Figure 3g with the etching angle α = $50^o$ while the piece shown on the right side of Figure 4a is the same one shown in Figure 3e, with the etching angle α = $30^o$. We can see that each piece is lighter on the upper side while darker on the lower side with gradual color change. The lighter side on the silicon piece in Figure 4a corresponds to the higher side shown in Figure 2a. All the SEM images in Figure 3 were taken on the black end of silicon pieces. To see what induces this gradual change of appearance in nanoscale, we take cross-sectional SEM images at different locations on the silicon piece, as shown in Figure 4b-g. We let the percentage stand for the location where the SEM is taken in the way from the light end to dark end. Figure 4b is taken at the light end. As we move from the light end to the dark end, the nanocones tend to be longer, shown in Figure 4b-g. Even though the nanocone length varies by different places, the slanted angle stays almost the same, about $20^o$ everywhere on this piece. We did not see the difference in the density of nanocones. A series of top view SEM images taken in a similar way demonstrate the uniform density better, shown in Figure. S2 in the supporting information.

Surface enhanced Raman spectroscopy (SERS) is a surface-sensitive technique that enhances Raman scattering by molecules adsorbed on rough metal surfaces. The enhancement factor can be as high as $10^{10}$ to enable single molecule detection.[34] The enhancement factor of is strongly dependent on the material and morphology of the rough metal surface. We already demonstrated that our straight cone black silicon deposited with 80 nm silver can enhance the Raman scattering of rhodamine 6g more than $10^7$ times and the fluorescence of rhodamine 6g by 30 times. In this article, our major purpose is to use SERS as a tool to characterize the various surface morphologies of the slanted nanocone black silicon. We will find the relationship between the SERS enhancement and surface morphology as well as other fabrication conditions. This will contribute to the design and optimization of SERS substrates.

First, we compare the SERS enhancement on sub-micron pyramids structure with and without nanocones. Since we can make nanocones on the inverse pyramids array structure (Figure 1c, d), which is the Klarite SERS substrate after gold being removed, we compare this structure with original Klarite SERS substrate for SERS. We call it black Klarite here, a photograph of which is shown in inset of Figure 1d. We also take the planar silicon with nanocone (planar black silicon) for SERS comparison. For the black Klarite and planar black silicon with the nanocones with height of 300 nm, we deposited 80 nm of gold by electron beam (e-beam) evaporation. Before the deposition of gold, 5 nm of Titanium was deposited as an adhesion layer between gold and silicon. The original Klarite SERS substrate has 300 nm thick gold on the surface.[33] In visible and near infrared range, the SERS enhancement of silver is usually higher than that of gold by two orders under the same nanostructure.[35] But silver will eventually get oxidized and lose enhancement. The reason we use gold instead of silver here is for fair comparison with original Klarite. Simple dropping and physical adsorption of analyte on surface will form non-uniform coverage as coffee stain effect. To get a uniform and quantitative characterization, a monolayer of the target molecule benzenethiol was formed on the gold surface by thiol-gold conjugation chemistry. The benzenethiol monolayer is formed by immersing the substrate in the solution of benzenethiol in ethanol with the concentration of 4mM for one hour.[36] Then we acquire the Raman spectra of benzenethiol by a Renishaw Raman system with the 785 nm laser with power of 1 mW and exposure time of 10 seconds.

Figure 5a shows the SERS spectra of benzenethiol on smooth gold surface (black curve), original Klarite SERS substrate (green curve) and planar black silicon (red curve) and black Klarite (blue curve). The characteristic Raman peaks of benzenethiol are marked out at the wavenumber of 695 cm$^{-1}$($\beta_{CCC}$ +$\nu_{CS}$), 1073 cm$^{-1}$($\beta_{CH}$), 1575 cm$^{-1}$($\nu_{CC}$); $\beta$ and $\nu$ indicate the in-plane bending and the stretching modes respectively.[37] In Figure 5a, the smooth gold surface hardly shows any Raman peaks while black Klarite and planar black silicon show higher peaks than original Klarite. For a quantitative analysis of SERS enhancement, we calculated the enhancement factors of each substrate based on the peak intensity at 1073 cm$^{-1}$ since all the Raman peaks are proportional in intensity on each substrate. As original Klarite substrate is proved to have enhancement factor of ~10$^6$,[33] we use it as a reference to compute the

enhancement factors for other substrates. The enhancement factor (EF) is calculated using the formula below:

$$EF = 10^6 \times \frac{I_{specimen}}{I_{Klarite}}, \qquad (1)$$

In which $10^6$ is the enhancement factor of original Klarite, $I_{specimen}$ and $I_{Klarite}$ are the Raman peak intensity at 1073 cm$^{-1}$ of the substrate of interest and original Klarite respectively. The calculated enhancement factors (EF) of different substrates are listed below:

**Table 1. The calculated enhancement factors of original Klarite SERS substrates, planar black silicon and black Klarite**

| Substrate | Original Klarite | Planar blackSi | Black Klarite |
|---|---|---|---|
| EF | $1 \times 10^6$ | $3.5 \times 10^6$ | $3.9 \times 10^6$ |

From Figure 5a and Table.1, we can see planar black silicon and black Klarite have larger SERS enhancement than original Klarite. The original Klarite is made with inverse pyramids pits for plasmon resonance at 785 nm to optimize SERS excited by this wavelength.[33] The EF of black Klarite and planar black silicon are $3.9 \times 10^6$ and $3.5 \times 10^6$ respectively, larger than the EF of original Klarite. Previously we got the enhancement factor of the order of $10^7$ by depositing 80 nm of silver on planar black silicon.[5] It is reasonable for gold to have lower enhancement factor than silver with the same nanostructure by 2 orders in visible and near-IR range. The result that black Klarite and planar black silicon have similar EF indicates that the micro-size inverse pyramids structure does not remarkably help the SERS of nanocone black silicon. Our explanation is that even though the nanocone forest creates more SERS hotspots for scattering light, at the same time it makes the reflection more diffusive thus destructs the plasmon resonance mode at 785 nm of the smooth inversed pyramids array. This explanation is simply verified by the appearance of the substrates. The black Klarite does not show the iridescence seen on original Klarite, shown in the inset of Figure 1d. The reason why EF of black Klarite is slighter higher than that of planar black silicon is probably only due to the larger surface area of inverted pyramids compared with planar surface. On the photograph in inset of Figure 1d, we can see the region of black inverse pyramids is darker than the surrounding regions of planar black silicon but with no iridescence color. The comparison of reflection spectra of black Klarite and original Klarite in the wavelength range from 650 nm to 850 nm in Figure 5b gives a more quantitative and convincing proof of our explanation. In Figure 5b we can see that the original Klarite shows a dip around 760 nm (close to 785 nm) while the planar black silicon and black Klarite do not show a dip there. But the reflection of black Klarite is lower than that of planar black silicon, which confirmed our observation on the inset of Figure 1d, that is, black Klarite is darker than planar black silicon. We improved the EF of Klarite SERS substrate by almost 4 times by making nanocones on inverse silicon pyramids array. Even if the improvement is within one order of magnitude, we only need to deposit 80 nm of gold, more than 2/3 thinner than 300 nm of gold on original Klairte.

Since SERS is strongly dependent on the size, structure and material of metal surfaces, we also use SERS to characterize the surface of the slanted nanocone black silicon. There are multiple factors that may affect SERS, including the type and thickness of metal being deposited, length, density and slanted angle of nanocones. For the purpose of SERS optimization, it is of great interest to investigate how the SERS enhancement factor depends on these factors. The first factor we want to investigate is the effect of slanted angle of nanocone.

To characterize the slanted angle dependence, we deposited 80nm of gold on to SNBS with different slanted angles, including the normal black silicon with straight up cones (zero slanted angle). A monolayer of benzenethiol was formed on the surface as the analyte for SERS. Then Raman spectra were taken at the dark end of SNBS because according to the results in Figure 4, the nanocone in this region has comparable length of 300nm to the straight up nanocone on vertical back silicon. Concerning the asymmetry of slanted nanocone, we need to consider polarization. Figure 6a,b are the schematics to show how are the propagation and polarization direction of laser excitation relative to the slanted direction of nanocone and the normal of substrate. Figure 6a shows the polarization parallel to the slanted direction while Figure 6b shows the polarization perpendicular to the slanted direction. We did not use polarizer for the collection of scattered light. Figure 6c shows the relation of Raman intensity at the 1073 cm$^{-1}$ peak along with its corresponding enhancement factor with different slanted angles, for both polarization directions. However, we did not see a clear monotonous trend of the Raman intensity with the slanted angle. But the polarization does matter for slanted nanocone. For normal black silicon with slanted angle of zero, the polarization direction makes no difference. For other slanted angles, the Raman intensity is always higher when the polarization is perpendicular to the slanted direction than when it is parallel. And this difference becomes more prominent as the increase of slanted angle.

To investigate the effect of cone length and metal thickness on SERS, we deposited gold with thickness of 30 nm and 80 nm onto a 30$^o$ SNBS for SERS. We have shown in Figure 4 there is a gradient in darkness, which is essentially a gradient in height of nanocone on the piece. Figure 7a shows enhancement factor from light end to dark end of a 30$^o$ SNBS with 30 nm gold for perpendicular and parallel polarizations. The percentage stands for the location where the spectrum is taken from the light end to dark end of the silicon piece (Figure 4a). For instance, 50% means the spectrum is taken when the laser spot is located halfway from the light end to dark end, 0% means at the edge of light end. From the spectra we can see the SERS signal intensity increases from the light end to dark end. SERS enhancement factor calculated based on the Raman peak at 1073 cm$^{-1}$ with equation (1) is indicated on the right vertical axis. Figure 7a shows that the SERS intensity increases from light end to dark end in an almost linear relationship for both polarizations. It also indicates there is no significant difference in SERS intensity at the same location for the two polarizations. However, the enhancement factor in this case is only around the order of 10$^4$ to 10$^5$, much weaker compared with that of 10$^6$~10$^7$ on black silicon deposited with 80 nm of gold. Figure 7b shows enhancement factor from light end to dark end of a 30$^o$ SNBS with 80 nm gold for perpendicular and parallel polarizations. With 80 nm gold deposited, the intensity-location relation is not monotonous as 30 nm gold sample in Figure 7a. The enhancement factors for both polarizations are of the order of $10^6$ to $10^7$ with the maximum enhancement factor around $7 \times 10^6$ except at the light end, where it is of the order of $10^5$. That means most places except the light end of SNBS with 80 nm of gold have comparable SERS enhancement with vertical nanocone planar black silicon with 80 nm of gold. Similar to the result shown in Figure 6, the enhancement factor for perpendicular polarization for SNBS with 80 nm of gold is always higher than that for parallel polarization at the same location.

To see what caused the difference in SERS results when SNBS is deposited with 30 nm and 80 nm of gold, we took top-view and cross-sectional SEM images of SNBS with 30 nm and 80 nm of gold, shown in Figure 8. Figure 8 a, e are top-view SEM images of SNBS with 30 nm gold and 80 nm gold respectively, where the arrow indicates the slanted direction of nanocone. After the gold being deposited on the nanocone, it forms particle like structure on the cone. By comparing Figure 8a and Figure 8e, we can see the feature size of 80 nm gold deposition is bigger than that of 30 nm gold deposition thus the spacing between adjacent particles is smaller. Closer spacing between particle creates stronger local electric field for stronger SERS enhancement as long as the particles are not touching, which is proven in literature.[38] Another explanation for stronger SERS on 80 nm gold samples is the redshift in the plasmonic band aligns more closely with the excitation wavelength (785 nm) providing higher enhancement than for smaller particles which do not show such a great redshift. Figure 8b-d are cross-sectional SEM images of SNBS deposited with 30 nm gold at the light end, halfway and dark end respectively. Figure 8f-h are cross-sectional SEM images for 80 nm gold deposition.

**Discussion**

We have demonstrated slanted nanocone produced on planar and micro-structured silicon and investigated its SERS properties. There are several questions need be answered.

What caused the difference in SERS results shown in Figure 7 when SNBS is deposited with 30 nm and 80 nm of gold? In the experimental results part, we already explained that 80 nm gold SNBS has stronger SERS because of stronger coupling and red shift of plasmonic band aligned closely with laser excitation. Why does SERS intensity increase monotonically as the cone length increases on the 30 nm gold sample but not on the 80 nm gold sample? Figure 8b-d show that as 30 nm gold deposition is so thin, there are lots of nanoparticles deposited on the slanted silicon nanocones. There is only particle plasmon but little surface plasmon along the slanted nanocones or the whole substrate surface. In this case the SERS intensity should be proportional to the number of nanoparticles which increase with the height of the nanocone. Figure 8f-h show that for the 80 nm gold sample, we start to have a continuous film covering the nanocone surface. In this case, the cone-cone plasmon coupling is the key to SERS. Previous results on polarized SERS on slanted silver nanorod array also demonstrated that perpendicular polarization gives stronger SERS due to stronger rod-rod coupling.[39] However, this is not exactly cone-cone coupling. Figure 8 e-h show that after 80 nm gold deposition, the cone is not fully covered by a uniform gold film. Actually most of gold stays on the top of cone like a bead. The local field enhancement for SERS mainly comes from the coupling between those gold beads or particles on top of the nanocones. So for the 80 nm gold SNBS, the enhancement is determined by the formation of those gold nanoparticles rather than by the silicon nanocones. That explains why the SERS intensity in 80 nm gold sample does not increase monotonically as that in case of 30 nm thick gold sample. The SERS intensity on the 80 nm gold SNBS is relatively uniform except at the light end, where the nanocones are too short for gold nanoparticles to form the particle-like shape as in the region with longer cones. Figure 6c shows that SERS intensity does not have a clear trend with slanted angles, which can also be explained with the formation of gold nanoparticles (described below). The enhancements on SNBS with different slanted angles for the same polarization are of the same order even though they are not identical. On Figure 6 and 7, at the same spot on an 80 nm gold SNBS, SERS is always stronger for perpendicular polarization compared to parallel polarization (with reference to slanted direction). On Figure 8e, we can see that in the slanted direction, the adjacent gold nanoparticles are further apart in the slanted direction compared to those in the perpendicular direction. With larger spacing between gold particles in the slanted direction, the coupling is weaker and plasmonic band is less aligned with laser excitation (785 nm). Therefore SERS is weaker. With larger slanted angle, the spacing between particles in the slanted direction is even larger. But the spacing between particles in the direction perpendicular to the slanted angle does not change. That explains why the difference in SERS between two polarizations becomes more prominent as the increase of slanted angle.

Besides, a more general question is, what is the additional contribution of the black silicon over the normal silicon nanostructure to the SERS enhancement factor? The major advantage of black silicon on SERS is its broadband and omnidirectional enhancement due to its irregular corrugated surface structure. For a normal silicon nanostructure, usually periodic structure, the coupling is highly wavelength and angle selective. So normal silicon nanostructure is usually iridescent. But black silicon looks black from all directions. That means black silicon can efficiently absorb light in very broad bandwidth from wide angles. Even with metal deposited (which suppose to give rise to a mirror surface), it still couples light from broad bandwidth and wide angles.[5] The absorption or coupling can be attributed to two factors. One is the gradient effective refractive index of the sharp nanocone layer. The other is diffraction of the irregular sub-wavelength nanocone array. In addition, after metal deposited, the sharpness of the nanocone helps create hot spots for SERS as "lightning rod" effect. Due to randomness in the structure, there is possibility of overwhelming interference (constructive) of surface plasmon at some location which will give rise to very high electromagnetic field ("hot spot"). The irregular corrugated nanostructure also provides additional surface plasmon coupled scattering path for the photons. All those factors contribute more to SERS than normal silicon nanostructure.

**Conclusion**

We demonstrated that the nanocone forest can be formed on a variety of silicon surfaces with 3D microstructures, including AFM cantilever tips, inverse pyramids array on commercial SERS substrate and positive pyramids on solar cell, with a 3-step self-masked reactive ion etching process. All these silicon surfaces become black after the treatment. SERS enhancement factor of $3.9\times10^6$ was achieved after depositing 80 nm of gold onto Klarite SERS substrate we made black, compared with that of $10^6$ of the original Klarite SERS substrate coated with 300 nm of gold. Slanted nanocone black silicon (SNBS) was produced with tilted etching process. SNBS deposited with 30 nm and 80 nm of gold shows the enhancement factor on the order of $10^4$~$10^5$ and $10^6$~$10^7$ respectively. The SERS intensity on SNBS with 30 nm of gold shows an almost linear dependence on the darkness or nanocone length but no dependence on the polarization of excitation light. While the SERS intensity on SNBS with 80 nm of gold shows no dependence on the darkness or nanocone length but shows dependence on the polarization of excitation light. The SERS intensity is stronger when the polarization is perpendicular to the slanted direction. We explain the SERS results with the formation of gold nanoparticles on the slanted silicon nanocones. The slanted nanocone black silicon integrated on 3D microstructures provides new dimensions for fabrication and optimization of SERS sensors as well as other nanophotonic sensors.

**Acknowledgement**

This work is in part supported by NSF grant ECCS 10-28568. The author would like thank Dr. Yaguang Lian for the assistance of plasma etching and Dr. Edmond Chow for the assistance of Raman spectroscopy measurement.

**Supporting Information Available**: More details on the materials and equipment used for experiments. Figure S1 is the SEM image of oxide nanomask formed on smooth silicon surface before the nanocones are etched out. Figure S2 is the topview SEM images of SNBS. Figure S3 shows raw Raman spectra of benzenethiol on SNBS coated by 30 nm and 80 nm of gold. This material is available free of charge via the Internet at http://pubs.acs.org.

**Figure Captions**

**Figure 1** (a) (scale bar = 5 μm) and (b) (scale bar = 2 μm) Nanocone forest made on silicon pyramids. (c) (scale bar = 2 μm) and (d) (scale bar = 1 μm) Nanocone forest made on inverted pyramids on silicon (black Klarite). (e) (scale bar = 10 μm) and (f) (scale bar = 1 μm) Nanocone forest made on silicon AFM tip. For each row, the SEM image in the right column is the magnified image of the region in cropped by the white square in the SEM image in the left column. The insets in the bottom left corners of (a), (c) and (e) are SEM images of silicon pyramids, original Klarite after gold being removed and silicon AFM tip before RIE treatment respectively. The scale bars in the insets in (a), (c) and (e) are 2 μm, 1 μm and 4 μm respectively. The insets in the upper right corners of (b), (d) and (f) are photographs to compare the appearances of silicon pyramids (solar cell), silicon inverted pyramids (Klairte) and AFM tip chips with (black) and without (original) nanocone forest. The scale bars in the insets in (b), (d) and (f) are 8 cm, 1 cm and 5 mm respectively.

**Figure 2** (a) Diagram of setup to produce slanted black silicon with RIE. (1 and 7) Electrodes to create an electric field (3) meant to accelerate ions (2) towards the surface of the titled silicon sample (4). One side (right side in this diagram) of the piece of silicon (4) is blocked up by stack of glass slides and the other side (left side in this diagram) is blocked with one glass slide. The thickness of silicon piece is exaggerated for illustration. The titled angle α is determined by the height of the stack of glass slide H and the distance between the two glass stacks L. tan(α)=H/L. (b) A photograph to show how is the silicon tilted and mounted on the carrier wafer. (c-e)

3-step $O_2$-$CHF_3$-$Ar+Cl_2$ fabrication process of SNBS. (c) A thin oxide layer (orange) formed on silicon surface (blue) by oxygen plasma. (d) Dispersed oxide nanomask formed by etching thin oxide layer with $CHF_3$ plasma. (e) Slanted nanocones etched by mixture plasma of $Cl_2$ and Ar (10:1).

**Figure 3** The dependence of nanocone slanted angle on etching tilted angle. Cross-section SEM of slanted nanocone black silicon when etching tilted angle of (a) 0° (b) 8° (c) 15° (d) 20° (e) 30° (f) 40° (g) 50° (h) 60° (i) =70°. (scale bar = 300 nm) (j) The plot of the angle dependence.

**Figure 4** (a) The gradient of darkness on two pieces of slanted nanocone black silicon with etching angle of 50° (piece on the left) and 30° (place on right). The lower side is lighter while the upper side is darker. (scale bar = 1 cm) (b-g) Cross-sectional SEM images taken on the SNBS on the right ($\alpha=30°$) in (a) to show the gradient of cone lengths from light end to dark end of SNBS. The percentage stands for the location where the SEM is taken in the way from the light end to dark end. L is the measured length of one cone. (b) 0%, no cone, L=0 nm. (c) 20%, L=166 nm. (d) 40%, L=204 nm. (e)60%, 232 nm. (f) 80%, 301 nm. (g) 100%, 386 nm. (scale bar = 300 nm)

**Figure 5** (a) Raman spectra of benzenethiol monolayer on different substrates including smooth gold surface (black), original Klarite SERS substrate (green), planar black silicon coated with 80 nm gold (red) and black Klarite substrate coated with 80 nm gold (blue). The exciting laser is with the wavelength of 785 nm, power of 1 mW and exposure time of 10 seconds. (a.u.) stands for arbitrary units. (b) Reflection spectra of different substrates including original Klarite SERS substrate (green), planar black silicon coated with 80 nm of gold (red) and black Klarite substrate coated with 80 nm of gold (blue). The smooth gold surface is regarded as 100% reflection mirror for reference.

**Figure 6**. Schematics to show the propagation direction and polarization of laser excitation relative to the slanted nanocones and the normal of substrate. (a) Polarization is parallel to the slanted direction. (b) Polarization is perpendicular to the slanted direction. S is Poynting vector or the propagation direction. E is direction of electric field or polarization. (c) shows the peak intensities at 1073 cm$^{-1}$ measured at the dark end of SNBS after 80 nm gold deposition with different slanted angle. The right vertical axis shows the enhancement factors calculated based on the peak intensities at 1073 cm$^{-1}$. The letter M stands for million or $\times 10^6$.

**Figure 7** (a) Enhancement factor calculated from the peak intensity at 1073 cm$^{-1}$ from the light end to dark end of SNBS with etching angle = 30$^o$ for 30 nm (a) and 80 nm (b) gold deposition respectively. The letter k stands for thousand or $\times 10^3$ and M stands for million or $\times 10^6$. Original Raman spectra are included in Figure S3 in supporting information.

**Figure 8** SEM images of 30$^o$ slanted nanocone black silicon deposited with gold of thickness of (a-d) 30 nm and (e-h) 80 nm. The top view SEM images of slanted nanocone black silicon deposited with (a) 30 nm of gold and (b) 80 nm of gold; the arrow indicates the slanted direction of nanocones. Cross-sectional SEM images of light end of the piece deposited with (b) 30 nm of gold and (f) 80 nm of gold, middle

in the piece deposited with (c) 30 nm of gold and (g) 80 nm of gold and the dark end of the piece with (d) 30 nm of gold and (h) 80 nm of gold. (scale bar = 300 nm)

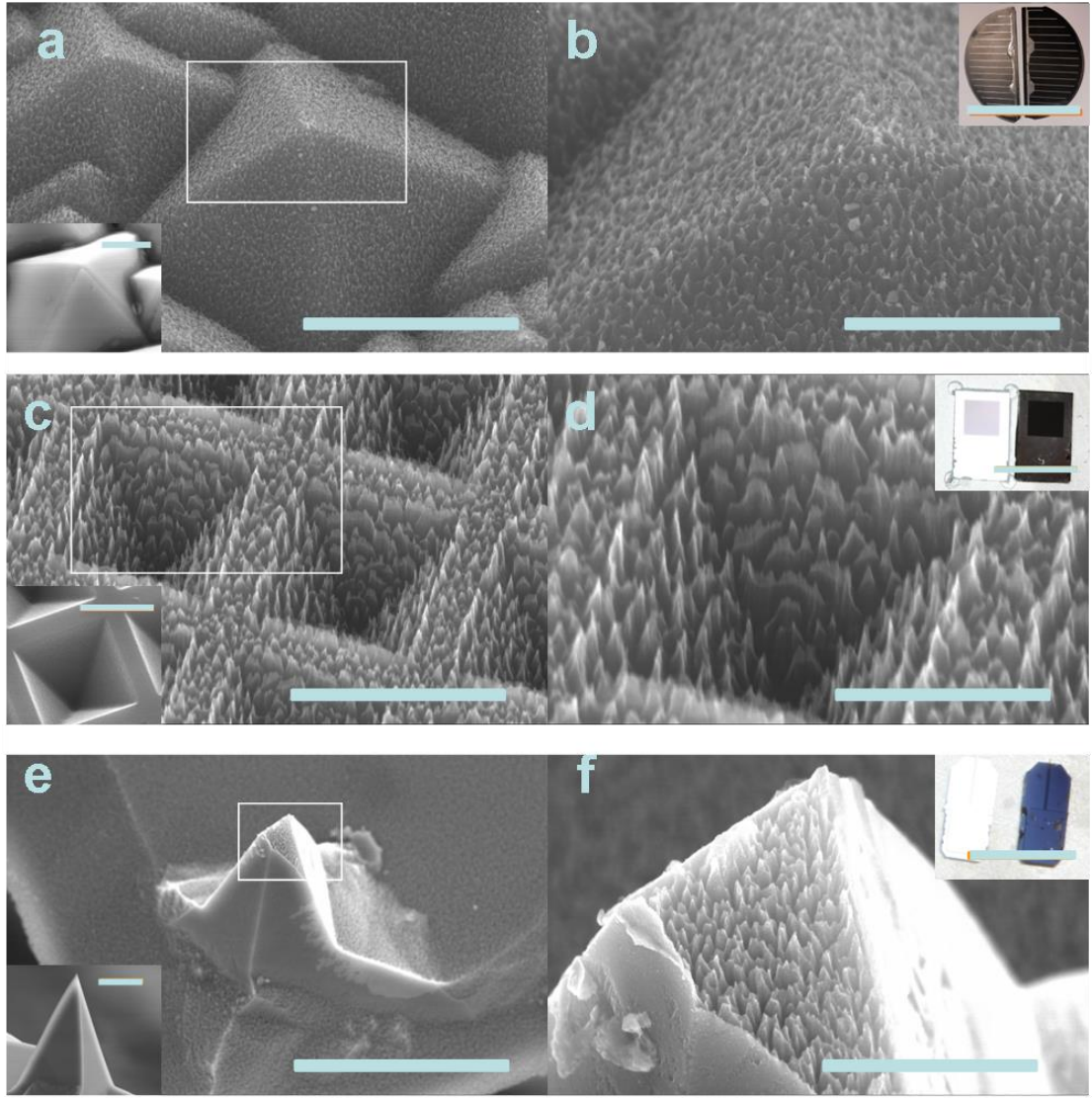

Figure 1

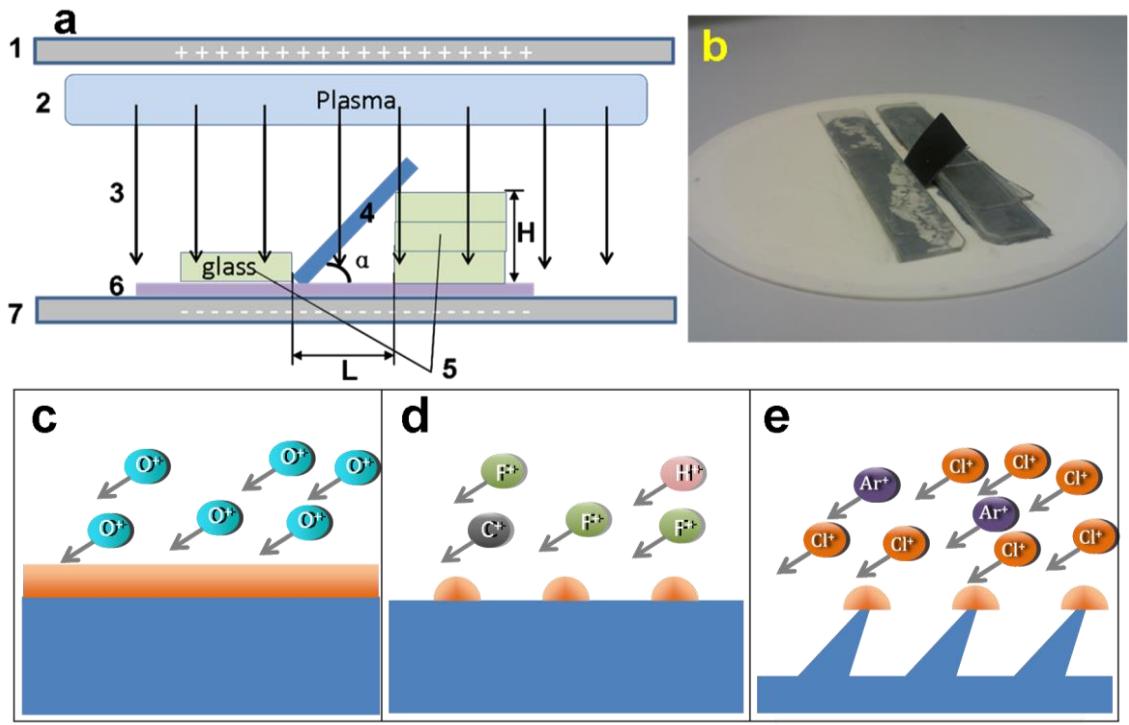

Figure 2

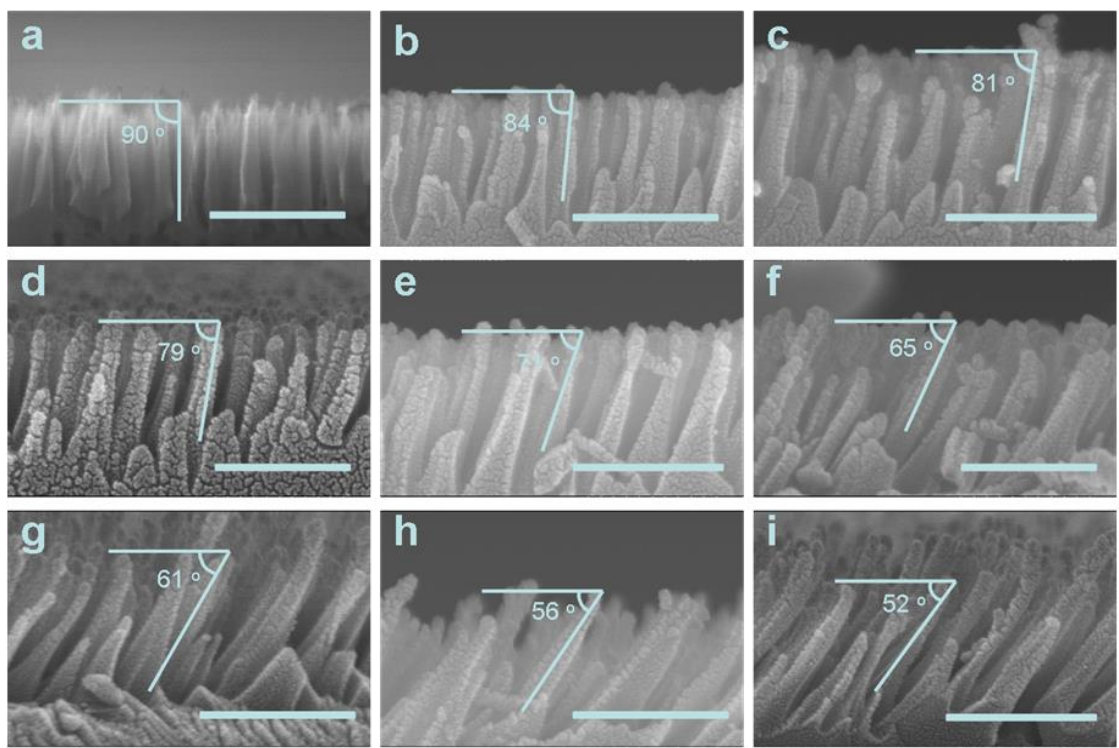

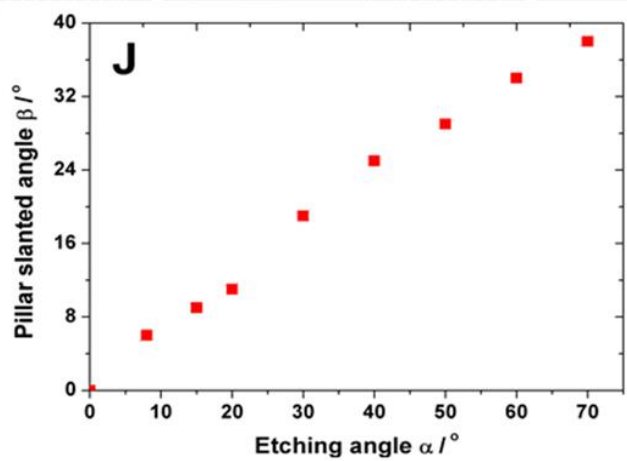

Figure 3

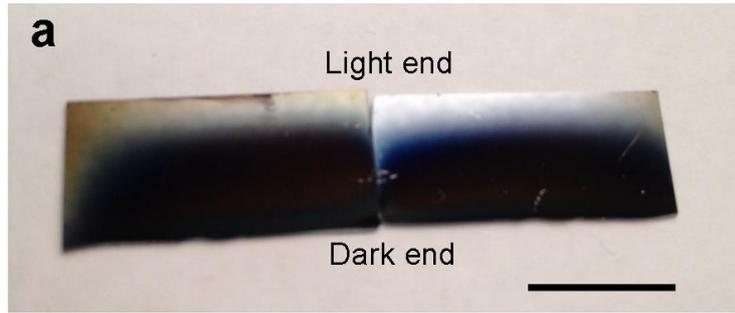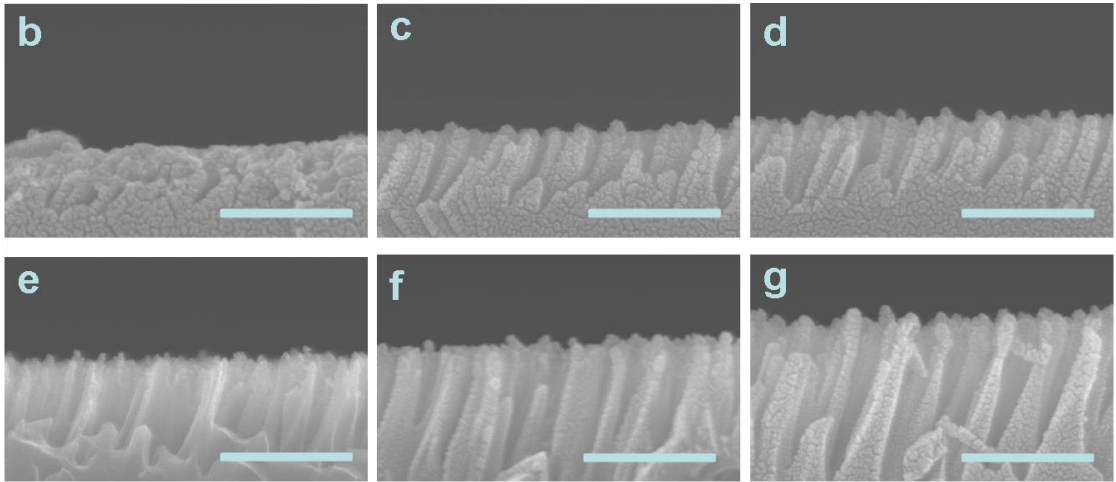

Figure 4

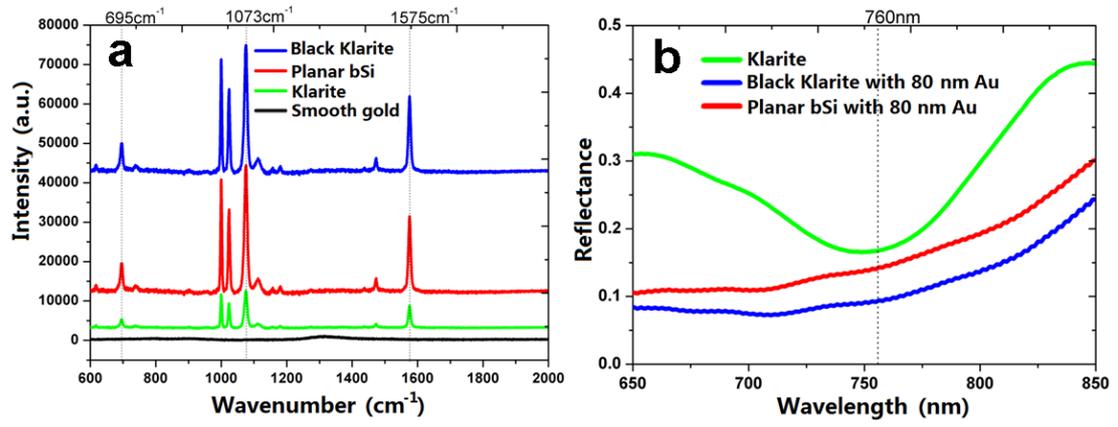

Figure 5

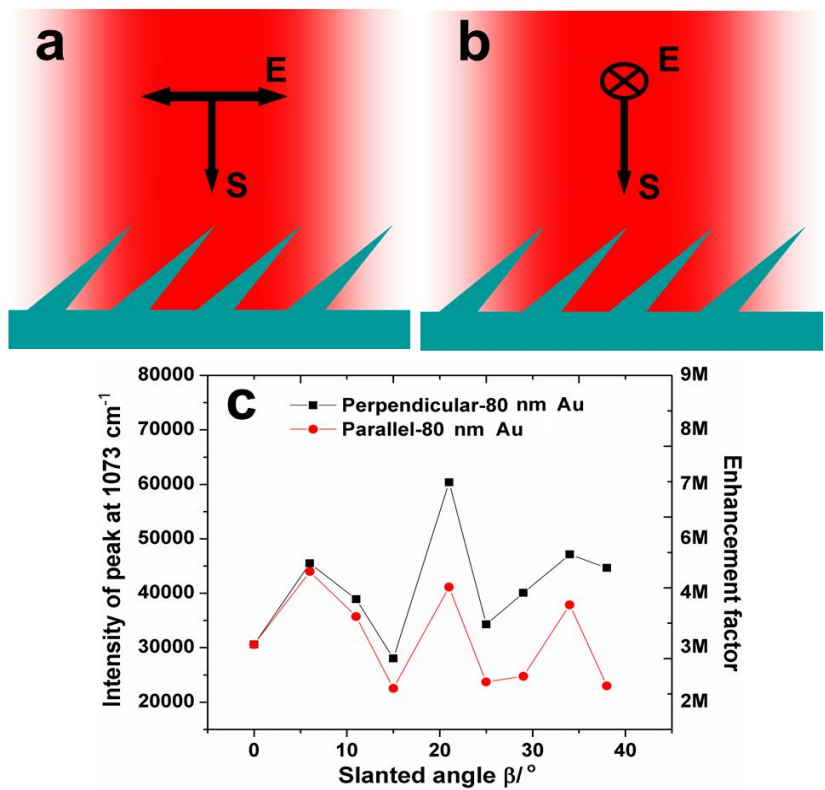

Figure 6

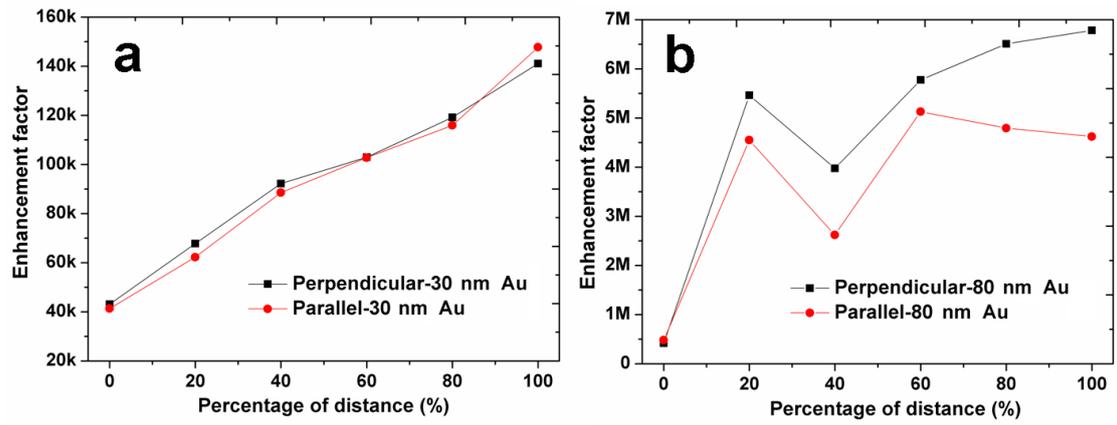

Figure 7

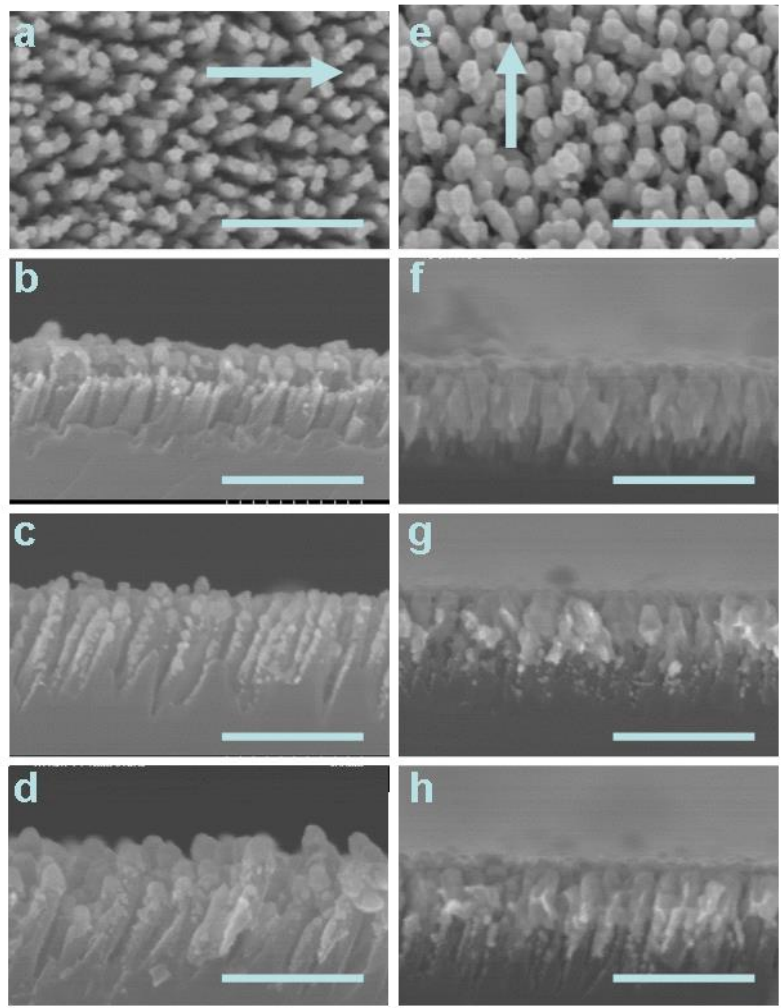

Figure 8

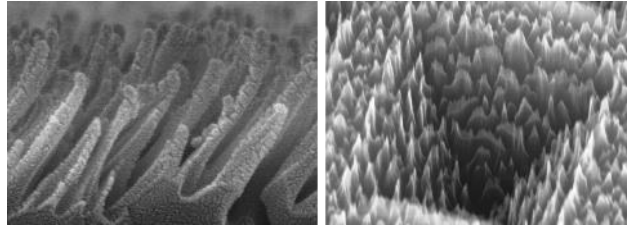

TOC image